\documentclass{aa}
\usepackage{hyperref}
\usepackage{graphicx}
\usepackage{amsmath}
\usepackage{cancel}

\graphicspath{{}}

\abstract{
    We investigate the ionic column density variability of the ionized outflows associated with NGC~7469, to estimate their location and power. This could allow a better understanding of
    galactic feedback of AGNs to their host galaxies. 
    Analysis of seven {\it XMM-Newton} grating observations from 2015 is reported.
    We use an individual-ion spectral fitting approach, and compare different epochs to accurately determine variability on time-scales of years, months, and days. We find no significant column density variability
    in a 10 year period
    implying that the outflow is far from the ionizing source. The implied lower bound on the ionization equilibrium time, 10 years, constrains the lower limit on the distance to be at least 12 pc, and up to 31 pc, much less but consistent with the 1 kpc wide starburst ring. The ionization distribution of column density is reconstructed from measured column densities, nicely matching results of two 2004 observations, with one large high ionization parameter ($\xi$) component at $2<\log \xi<3.5$, and one at $0.5<\log \xi<1$ in cgs units. 
    The strong dependence of the expression for kinetic power, $\propto1/\xi$, hampers tight constraints on the feedback mechanism of outflows with a large range in ionization parameter, which is often observed and indicates a non-conical outflow.
    The kinetic power of the outflow is estimated here to be within 0.4 and 60 \% of the Eddington luminosity, depending on the ion used to estimate $\xi$.
}

\title{Multi-wavelength campaign on NGC7469 II. \\Column densities and variability in the X-ray spectrum}

\author{U. Peretz\inst{1} \and E. Behar\inst{1} \and G. A. Kriss\inst{2} \and J. Kaastra\inst{3,4} \and N. Arav\inst{1,5} \and S. Bianchi\inst{6} \and G. Branduardi-Raymont\inst{7} \and M. Cappi\inst{8} \and E. Costantini\inst{3} \and B. De Marco\inst{9} \and L. Di Gesu\inst{10} \and J. Ebrero\inst{11} \and S. Kaspi\inst{12} \and M. Mehdipour\inst{3} \and R. Middei\inst{6} \and  S. Paltani\inst{10} \and P.O. Petrucci\inst{13} \and G. Ponti\inst{14} \and F. Ursini\inst{8}}

\institute{Department of Physics, Technion, Haifa 32000, Israel \and Space Telescope Science Institute, 3700 San Martin Drive, Baltimore, MD 21218, USA \and SRON Netherlands Institute for Space Research, Sorbonnelaan 2, 3584 CA Utrecht, The Netherlands \and Leiden Observatory, Leiden University, PO Box 9513, 2300 RA Leiden, The Netherlands \and Department of Physics, Virginia Tech, Blacksburg, VA 24061, USA \and Dipartimento di Matematica e Fisica, Universit\`{a} degli Studi Roma Tre, via della Vasca Navale 84, 00146 Roma, Italy \and Mullard Space Science Laboratory, University College London, Holmbury St. Mary, Dorking, Surrey, RH5 6NT, UK \and INAF-IASF Bologna, via Gobetti 101, 40129 Bologna, Italy \and Nicolaus Copernicus Astronomical Center, Polish Academy of Sciences, Bartycka 18, PL-00-716 Warsaw, Poland \and Department of Astronomy, University of Geneva, 16 Chemin d'Ecogia, 1290 Versoix, Switzerland \and European Space Astronomy Centre, PO Box 78, 28691 Villanueva de la Ca\~{n}ada, Madrid, Spain \and School of Physics and Astronomy, Tel Aviv University, Tel Aviv 69978, Israel \and Univ. Grenoble Alpes, CNRS, IPAG, F-38000 Grenoble, France \and Max-Planck-Institut f\"{u}r extraterrestrische Physik, Giessenbachstrasse, 85748 Garching, Germany}

\begin{document}
    \titlerunning{Column densities and variability in the X-ray spectrum}
    \maketitle    
        
    \section{Introduction}
    Active galactic nuclei (AGN) are the most persistent luminous objects in the universe. Observed in all wavelengths from Radio to X-rays, they are powered by accretion of matter on to a super massive black hole.
    
    Among the plethora of phenomenon they exhibit, 50\% of type 1 AGN feature ionized outflows. The launching mechanism of these winds remains in debate, and suggestions vary from thermal evaporation \citep{Krolik01} to line driving \citep{Proga00} and magnetic hydrodynamics \citep{Fukumura10}.    
    These AGN winds are observed in a multitude of absorption lines of different ions, in both UV and X-rays \citep{Crenshaw03a}. These lines are  ubiquitously blueshifted with respect to the rest-frame of the host galaxy, with velocities often consistent between the X-rays and the UV \citep[e.g.][]{Gabel03a}, suggesting they are part of the same kinematic structure.
    
    If these outflows are indeed associated with the AGN, an important question is whether the energy or mass they deposit is important for galactic 
    evolution by means of energy feedback. The kinetic power of these outflows scales with $v^3$, the outflow velocity, which is
    typically a few 100 km s$^{-1}$ \citep{Kaastra02}. These low velocities limit the efficiency of these outflows as a feedback mechanism.
    However, some outflows feature velocities of a few 1000 km s$^{-1}$, NGC 7469 for example exhibits a fast component at a blueshift of 2000 km s$^{-1}$.
    
    AGN winds have been the focus of studies relating change in absorption troughs in AGN spectra to the distance and density of the associated
    outflows. Examples in both X-rays and UV analysis can be found in \citet{Behar03,Gabel05var,Kaastra12,Arav15,Ebrero16,Costantini16}. \citet{Arav15} for example constrain the distance of the outflow in NGC 5548 to be at least a few pc from the AGN source, with distances up to more than 100 pc. These large distances lead to an 
    ambiguity of whether the AGN is responsible for driving these outflows directly.
    
    This is the second paper as part of a multi-wavelength observation campaign on NGC 7469. \citet{Behar17} derived outflow parameters using global fit 
    models of photo-ionized plasmas. 
    We continue the examination of the {\it XMM-Newton} RGS spectrum focusing on measurement of the column densities. 
    In addition to the seven observations observed on a logarithmic timescale during the 2015 campaign, we analyze archival data from 2004.
    With these data we compare changes on timescales of years, months and days, 
    with the intent of seeking variability in absorption troughs, 
    and through this to constrain the distance of the outflow from the AGN. This, along with a measurement of the kinetic power of the outflow
    will determine the role the outflow plays in coupling the AGN to its host galaxy.
    
    \section{Data}
    \begin{figure*}
        \centering
    	\includegraphics[trim=0.5cm 0 5.0cm 0,clip,width=0.9\linewidth]{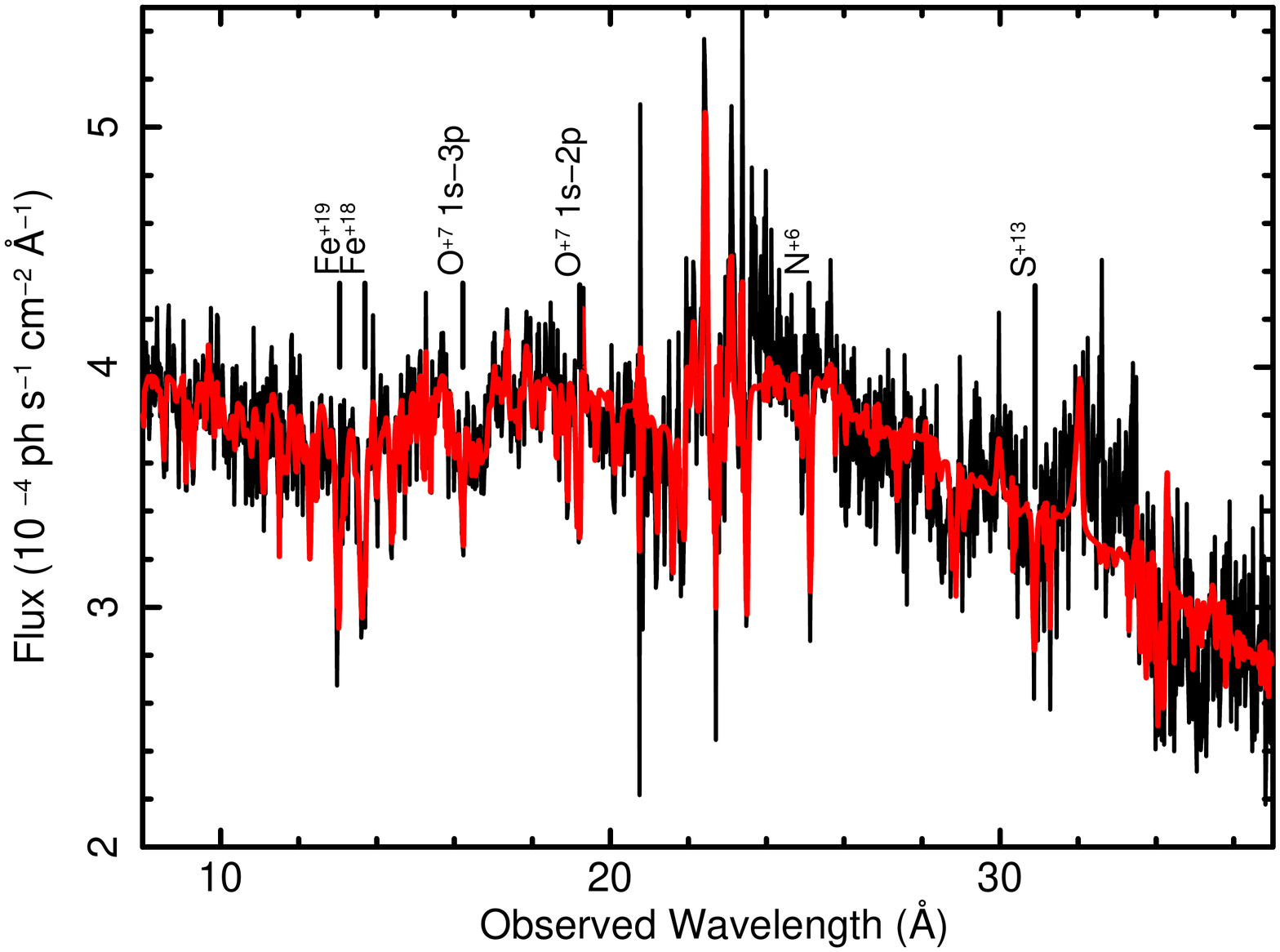}
    	\caption{The NGC 7469 2015 combined RGS spectrum (black) and best-fit model (red), with some prominent absorption lines marked. Data are re-binned to 60 m\AA\ in the image. Emission and local absorption are taken from \citet{Behar17}, and fitting is done on the column densities of individual ions. The unidentified features missing in the fit, for example emission at 33\AA\ and absorption at 28\AA, are discussed in \citet{Behar17}.}\label{fig:spectrum}
    \end{figure*}
    
    {\it XMM-Newton} observed NGC~7469 as part of the multi-wavelength campaign 7 times during 2015 for a total duration of 640 ks. 
    The observation log is shown in Table \ref{tab:log}, including previous observations published in \citet{Blustin07}. We use the RGS (1 and 2) data from
    all observations to constrain variability in absorption troughs.
    The RGS spectra are reduced using `rgsproc' within the software package SAS 15\footnote{\url{http://xmm-tools.cosmos.esa.int}} and combined using the standard RGS command, `rgscombine'. The reduction is detailed in \citet{Behar17}.
    The spectral fitting in the present paper is done on grouped spectra, re-binned to 20 m\AA\ (grouping two default SAS bins).
    The full 2015 RGS spectrum (black) and best-fit model (red) are shown in Fig. \ref{fig:spectrum}, and the model is described in Sec. \ref{sec:method}.
    
    \begin{table}
    	\centering
    	\caption{Observation log}\label{tab:log}
    	\begin{tabular}{c|c|c|c|c}
    		& obs. Id & start date & RGS & exposure\\
    		&&& $10^5$ cts & ks \\
    		\hline
    		a & 0207090101 & 2004-Nov-30 & 1.42 & 84.7\\
    		b & 0207090201 & 2004-Dec-03 & 1.13 & 78.8\\
    		1 & 0760350201 & 2015-Jun-12 & 1.36 & 89.5\\
    		2 & 0760350301 & 2015-Nov-24 & 1.41 & 85.6\\
    		3 & 0760350401 & 2015-Dec-15 & 1.18 & 84.0\\
    		4 & 0760350501 & 2015-Dec-23 & 0.97 & 89.5\\
    		5 & 0760350601 & 2015-Dec-24 & 1.04 & 91.5\\
    		6 & 0760350701 & 2015-Dec-26 & 1.19 & 96.7\\
    		7 & 0760350801 & 2015-Dec-28 & 1.23 & 100.2\\	
    	\end{tabular}
    \end{table}
    
\begin{figure}
    	\centering
    	\includegraphics[trim=1.7cm 0 2cm 1.9cm,clip,width=\linewidth]{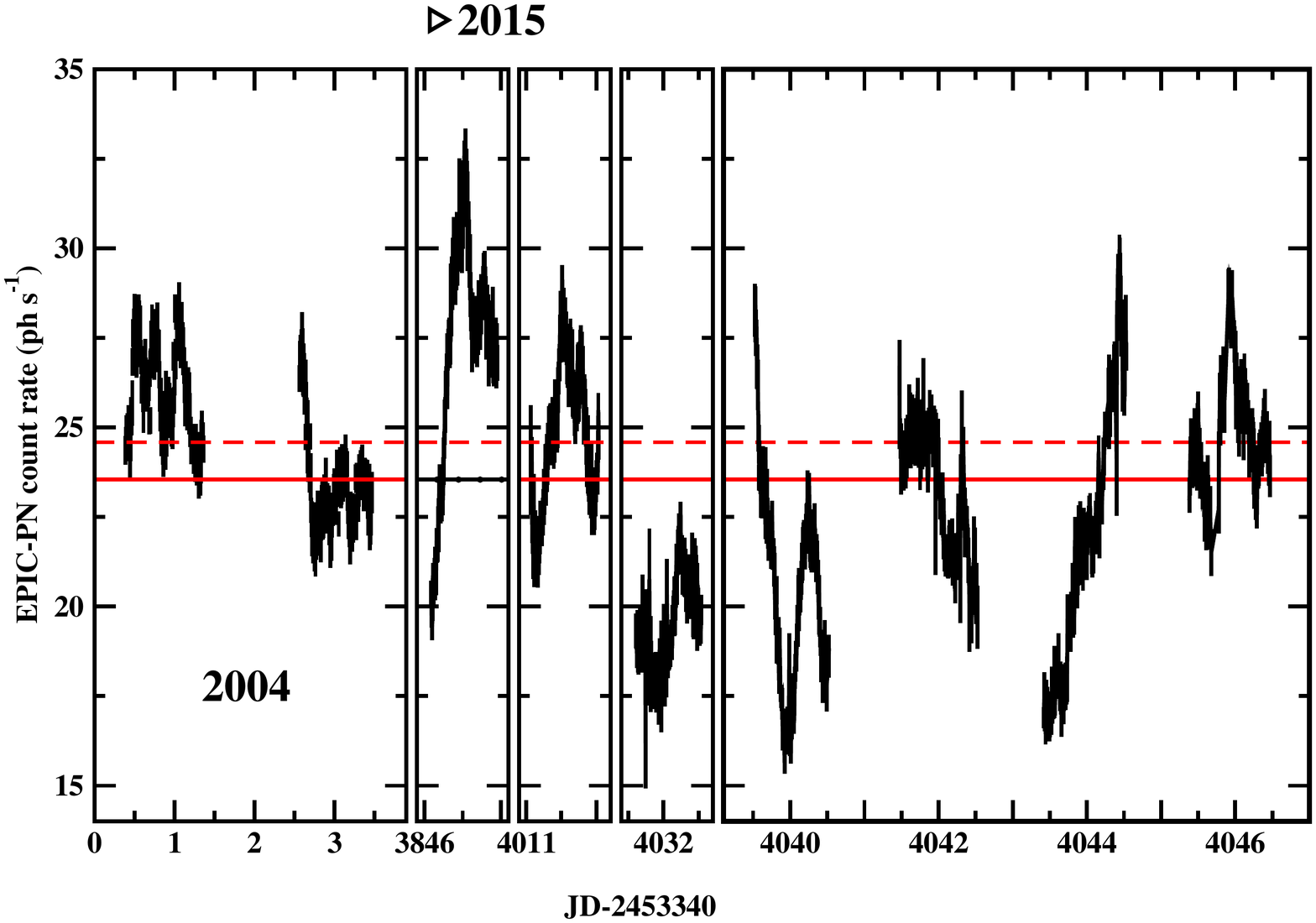}
    	\caption{EPIC-pn lightcurves (0.3-10 keV), 2 observations during 2004 and the 7 from the 2015 campaign.
    		While on a time scale of a day the flux from NGC~7469 changes by a factor of 2, overall the average flux remains constant - marked by the solid red line. In dashed red is the line defining high and low state, $0,4\sigma$ above the average.}\label{fig:pnlc}
    \end{figure}

    The EPIC-pn lightcurve of NGC~7469 is presented in Fig.~\ref{fig:pnlc}.
    An interesting feature  is the rapid change of photon flux on an hourly basis, while the average seems to remain constant over years.  
	The mean EPIC-pn count rate (count s$^{-1}$) for the 2004 observations is 24.7, with a standard deviation of $\sigma=1.9$, and for 2015 the mean is 23.2 with $\sigma=3.5$. 
    
    \section{Spectral modeling}
    
    \subsection{Method}\label{sec:method}
    We first model the 2015 combined spectrum  since
    column densities between observations in the campaign are consistent within 90\% uncertainties (See Sec. \ref{sec:results}).
    This agreement between the different observations, within the larger uncertainties of individual observations, is a clear indication in favor of using the combined spectrum, at least initially. All uncertainties we quote in this paper are 90\% confidence intervals.
    
    Following the ion-by-ion fitting approach by \citet{Holczer07}, 
    we fit the continuum $I_0$ along with the ionic column densities, $N_i$, which are this paper's main goal. The transmission equation is given by:
    \begin{equation}\label{eq:intensity}
    I(\lambda)/I_0=1-\left(1-\hspace{0.05cm}\mathrm{\textbf{e}}^{-\sum_i N_i\sigma_i(\lambda)}\right)C
    \end{equation}
    where $I(\lambda)$ is the observed continuum intensity, $I_0$ is the unabsorbed continuum intensity, $\sigma_i$ is the absorption cross section depending on photon energy. 
    The covering 
    fraction is $C$, with $C=0$ indicating no absorption and $C=1$ indicating the source is entirely covered by the outflow. Some results in the UV suggest the covering fraction is ion dependent or even velocity dependent  \citep[e.g][]{Arav12}, but the much smaller X-ray source is not expected to be partially covered.    
    The X-ray continuum of NGC~7469 in the RGS band can be modeled by a single powerlaw. A complete X-ray continuum model based on the EPIC spectra will be presented by Middei et al. (in preparation).
    The powerlaw is given by:
    \begin{equation}\label{eq:cont}
       I_0(E)=A\left(\frac E {\mathrm{1keV}}\right)^{-\Gamma}
    \end{equation}
    with the norm $A$ and the slope $\Gamma$ as free parameters.
        
    On top of the absorbed continuum $I(\lambda)$ we observe emission lines.  These lines were modeled by \citet{Behar17}, and include both photo-driven and collisionally excited lines. They are fixed in our model and are assumed not to be absorbed by the outflow.
    
    The absorption cross section is given by
    \begin{align}
    \sigma_i(\lambda)=&\sigma_i^\mathrm{edge}(\lambda)+\sigma_i^\mathrm{line}(\lambda)=\\
    =&\sigma_i^\mathrm{edge}(\lambda)+(\pi e^2/m_e c)\sum_{j<k} f_{jk} \phi (\lambda-\lambda_{jk})
    \end{align}
    Here $\sigma_i^\mathrm{edge}$ describes the ionization edge of ion $i$, $\phi(\lambda)$ is the Voigt line profile and the sum is over all the strong ion line transitions $j\rightarrow k$, $e$ is the electron charge, $m_e$ the electron mass, and $f_{jk}$ are the oscillator strengths. 
    All transitions are assumed to be from the ground level.
    We use the oscillator strengths and ionization edges calculated using the HULLAC  atomic code \citep{BarShalom01} as used in \citet{Holczer07}.
    
    The parameters determining the profile shape and position $\phi$ are ion temperature, turbulent velocity, and outflow velocity. 
    The temperature and turbulent velocity broadenings seen in the UV \citep{Scott05} are below the  RGS resolution of $\Delta\lambda\approx70$m\AA.
    Thus, in order to constrain simultaneously the covering factor, the turbulent velocity, and the ion column density one needs 3 measurable lines of a given ion (See Eq. \ref{eq:intensity}).
    N$^{+6}$ is the best ion providing 3 lines unambiguously visible in the spectrum. These are observed at wavelengths of 25.18\AA, 21.25\AA, and 20.15\AA.  Nonetheless, the best fit favors a covering factor of 1.0 with the 90\% confidence interval ranging down to 0.8 when all ions are taken into account. The uncertainty in the continuum adds another level of uncertainty here, so we make no claims regarding covering factor and hold it frozen to 1.0. 
    
\begin{figure}
        \includegraphics[trim=0 0 4cm 2cm,clip,width=\linewidth]{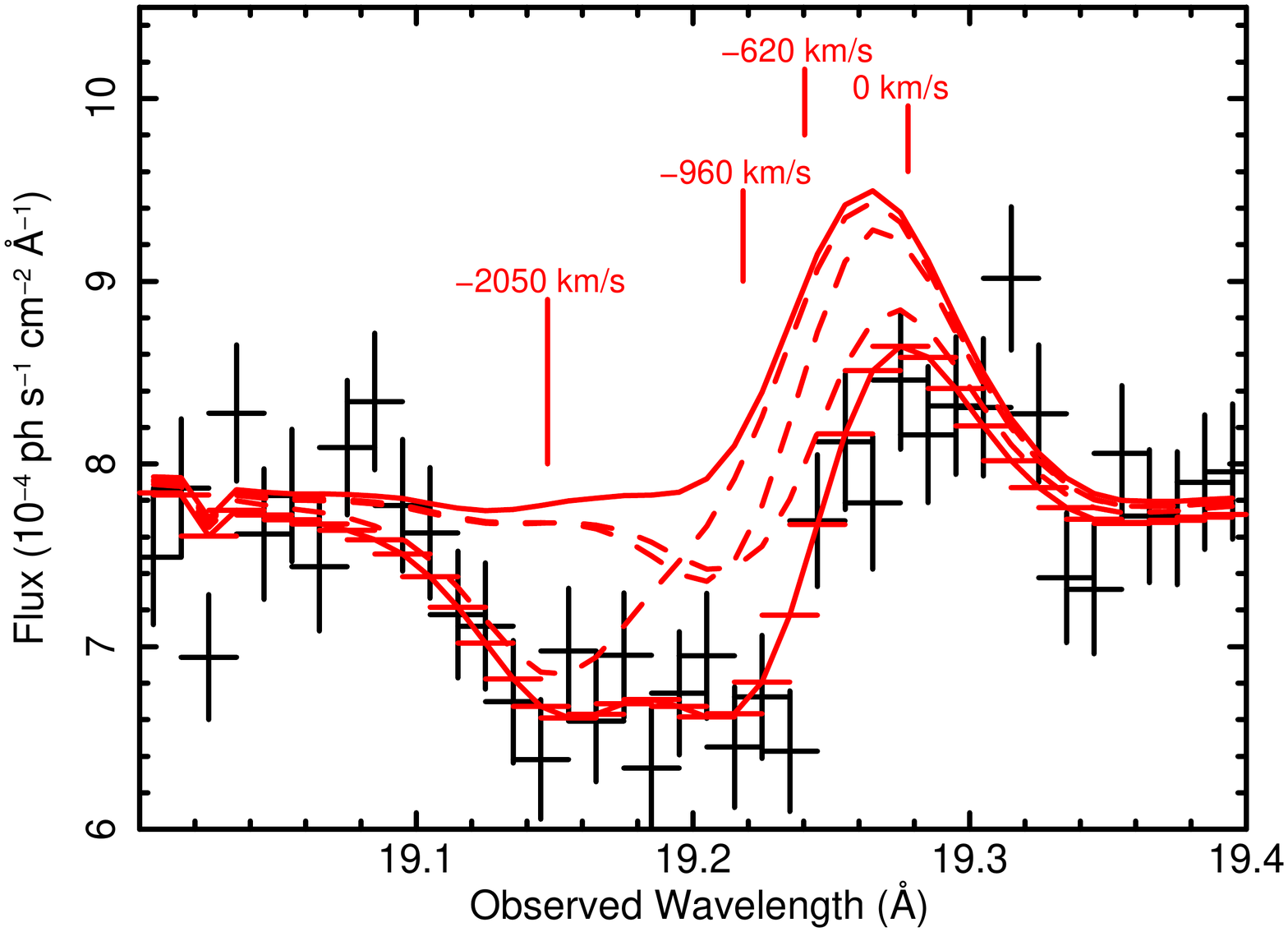}
        \caption{Break down of velocity components in the O$^{+7}$ line. 
            The unabsorbed flux is the upper line, and absorption is marked with dashed lines. 
            The bottom line is the full model. We apply absorption only to the continuum, and not the emission. The fast (left most) and middle velocities correspond to the centroids of the double trough in the model, while the slow component absorbs around the blue-shifted (un-absorbed) emission line.}\label{fig:o7}
    \end{figure}

    Since constraining the line profile parameters is not the goal of this paper, we fix the ion temperature at 0.1 keV. 
    We then fit only the O$^{+7}$ Ly$\alpha$ doublet line at  the observed wavelength of 19.2\AA\ with the outflow and turbulent velocities thawed and set initially to the values of \citet{Behar17} in order to determine 
    them. 
    Fig. \ref{fig:o7} shows the contribution of each velocity component to the absorption profile.
    The fit favors a 3 velocity model over 2 in accordance with these two papers, decreasing reduced $\chi^2$ ($\chi^2/$d.o.f.) by 0.5 from the 2-component to the 3-component model.
    The best-fit three components have velocities of -620, -960, and -2050 km s$^{-1}$ and turbulent velocities of 80, 40, 50 km s$^{-1}$ respectively. 
    Three components are also favored by \citet{Scott05} and \citet{Behar17}.
    Though the fit converges we are not able to obtain meaningful uncertainties on these parameters. 
    We leave them frozen for the rest of the fit, freeing them for one final iteration after the ion column densities are constrained.
    
    The fitted model parameters are thus the powerlaw normalization, the powerlaw slope, and the column density per ion. In addition,
    the three outflow velocities and three turbulent velocities are constrained once at the beginning according to O$^{+7}$, and one more time at the end.
    The strength of this model\footnote{The code for the model can be found in \url{https://github.com/uperetz/AstroTools}, including a full graphical suite for fitting models to fits files. The README details the contents of the directory.} lies in the independence of the ionic free parameters.     
        
    \subsection{Column densities}\label{sec:results}
    
     The full 2015 spectrum (black) and best-fit model (red) is seen in Fig. \ref{fig:spectrum}, with a best-fit reduced $\chi^2$ of 1.4.
     For the 2004 spectra we obtain a  reduced $\chi^2$ of $1.28$. There are 1450 spectral bins and 64 free parameters.
     We also re-measure column densities from the 2004 spectra previously done by \cite{Blustin07}. This is done in order to maintain consistency in the comparison with the 2015 spectra using the same code and atomic data.
     \citet{Blustin07} finds 2 velocities, but we retain the 3 velocity model for a consistent comparison with 2015. There is no increase of reduced $\chi^2$ compared to the two velocity model, suggesting the kinematics remain similar  over a timescale of years.
     In Table \ref{table:continuum} the continuum parameters of both epochs, 2004 and 2015, are presented.
    
    \begin{table}
        \centering
        \caption{Unabsorbed continuum parameters (Eq. \ref{eq:cont})}\label{table:continuum}
        \begin{tabular}[b]{c|cc}
            & 2015   & 2004                  \\
            \hline
            $\Gamma$ & $2.129\pm0.005$ & $2.335\pm0.009$ \\
            $A$ $^1$    & $11.15\pm0.03$ & $13.75\pm0.07$   \\
            Flux $^2$ & $2.8\pm0.2$    & $3.8\pm0.5$   \\
            \multicolumn{3}{l}{\small $^1$ $10^{-3}$ ph keV$^{-1}$ s$^{-1}$ cm$^{-2}$} \\
            \multicolumn{3}{l}{\small $^2$ $10^{-11}$ erg s$^{-1}$ cm$^{-2}$, RGS band 0.3-1.5 keV (8-37\AA)} \\
        \end{tabular}
    \end{table}
    
    \begin{table*}
    	\centering
    	\caption{Summed ionic column densities}\label{tab:cols}
    	\begin{minipage}[t][][b]{0.49\linewidth}
    		\centering
    		\begin{tabular}[b]{c|ccc}
    			& 2015                  & 2004                   & \citet{Blustin07}       \\
    			Ion     & $10^{15}$ cm$^{-2}$   & $10^{15}$ cm$^{-2}  $  & $10^{15}$ cm$^{-2}$ \\
    			\hline
    			C$^{+4}$   & $9.6   _{-6.0  }^{+8.1   }$ & $<6.0                     $ & ...                     \\
    			C$^{+5}$   & $70    _{-30 }^{+40    }$ & $60    _{-30 }^{+40    }$ & $50    _{-30 }^{+30    }$ \\
    			N$^{+5}$   & $2.0   _{-2.0  }^{+3.8   }$ & $0.9   _{-0.9  }^{+3.3   }$ & $2.5   _{-2.4  }^{+3.8   }$ \\
    			N$^{+6}$   & $100   _{-30 }^{+40    }$ & $20    _{-10 }^{+50    }$ & $30    _{-10 }^{+20    }$ \\
    			O$^{+4}$   & $30    _{-20 }^{+30    }$ & $<4.1                     $ & ...                     \\
    			O$^{+5}$   & $<1.3                     $ & $<1.6                     $ & ...                     \\
    			O$^{+6}$   & $50    _{-20 }^{+30    }$ & $200   _{-39.7 }^{+50    }$ & $20    _{-10 }^{+10    }$ \\
    			Ne$^{+8}$  & $90    _{-50 }^{+80    }$ & $200   _{-70 }^{+70    }$ & $200   _{-80 }^{+90    }$ \\
    			Ne$^{+9}$  & $70    _{-50 }^{+90    }$ & $200   _{-180}^{+300   }$ & $200   _{-120}^{+200   }$ \\
    			Mg$^{+10}$ & $1700  _{-740}^{+1100  }$ & $1000  _{-920}^{+3800  }$ & $30    _{-20 }^{+70    }$ \\
    			Mg$^{+11}$ & $40    _{-30 }^{+80    }$ & $10    _{-10 }^{+1600  }$ & $30    _{-30 }^{+70    }$ \\
    			S$^{+12}$  & $<0.4                     $ & $50    _{-30 }^{+300   }$ & $5.0   _{-3.8  }^{+5.0   }$ \\
    			S$^{+13}$  & $10    _{-4.9  }^{+7.0   }$ & $0.7   _{-0.7  }^{+5.2   }$ & ...                     \\        
    		\end{tabular}                                          
    	\end{minipage}
    	\begin{minipage}[t][][b]{0.49\linewidth}
    		\centering
    		\begin{tabular}{c|ccc}
    			& 2015                  & 2004                   & \citet{Blustin07}       \\
    			Ion     & $10^{15}$ cm$^{-2}$   & $10^{15}$ cm$^{-2}   $ & $10^{15}$ cm$^{-2} $\\
    			\hline                                                   
    			Fe$^{+3}$  & $2.2   _{-2.2  }^{+2.9   }$ & $<4.7                     $ & $<0.0                     $ \\
    			Fe$^{+4}$  & $<2.4                     $ & $1.3   _{-1.3  }^{+4.3   }$ & ...                     \\
    			Fe$^{+5}$  & $3.4   _{-1.8  }^{+1.8   }$ & $<3.7                     $ & ...                     \\
    			Fe$^{+6}$  & $<1.0                     $ & $1.3   _{-1.3  }^{+2.7   }$ & ...                     \\
    			Fe$^{+7}$  & $2.2   _{-1.3  }^{+1.3   }$ & $<1.5                     $ & ...                     \\
    			Fe$^{+8}$  & $2.5   _{-1.7  }^{+1.9   }$ & $2.7   _{-1.7  }^{+2.1   }$ & $5.0   _{-2.5  }^{+2.9   }$ \\
    			Fe$^{+9}$  & $2.7   _{-1.1  }^{+3.3   }$ & $1.2   _{-1.2  }^{+2.0   }$ & $6.3   _{-2.3  }^{+1.6   }$ \\
    			Fe$^{+10}$ & $1.3   _{-1.0  }^{+1.1   }$ & $1.2   _{-1.2  }^{+2.1   }$ & $3.2   _{-2.5  }^{+3.1   }$ \\
    			Fe$^{+11}$ & $2.8   _{-1.3  }^{+1.3   }$ & $3.5   _{-2.1  }^{+2.2   }$ & $1.0   _{-1.0  }^{+2.2   }$ \\
    			Fe$^{+12}$ & $3.2   _{-1.8  }^{+1.9   }$ & $3.1   _{-1.8  }^{+2.0   }$ & $5.0   _{-1.8  }^{+2.9   }$ \\
    			Fe$^{+13}$ & $0.3   _{-0.3  }^{+1.0   }$ & $0.7   _{-0.7  }^{+1.9   }$ & $2.5   _{-1.9  }^{+2.5   }$ \\
    			Fe$^{+14}$ & $1.0   _{-0.7  }^{+1.0   }$ & $<0.8                     $ & $2.0   _{-1.9  }^{+3.0   }$ \\
    			Fe$^{+15}$ & $0.1   _{-0.1  }^{+1.7   }$ & $1.4   _{-1.4  }^{+2.7   }$ & $2.0   _{-1.8  }^{+2.0   }$ \\
    			Fe$^{+16}$ & $10    _{-6.6  }^{+9.0   }$ & $40    _{-10 }^{+20    }$ & $10.0  _{-5.0  }^{+5.8   }$ \\
    			Fe$^{+17}$ & $20    _{-5 }^{+7   }$ & $40    _{-10  }^{+10    }$ & $20    _{-4  }^{+5   }$ \\
    			Fe$^{+18}$ & $30    _{-5  }^{+10   }$ & $30    _{-15 }^{+20    }$ & $30    _{-5  }^{+6   }$ \\
    			Fe$^{+19}$ & $30    _{-12 }^{+10    }$ & $20    _{-10 }^{+10    }$ & $20    _{-4  }^{+5   }$ \\
    			Fe$^{+20}$ & $20    _{-12 }^{+20    }$ & $80    _{-50 }^{+100   }$ & $40    _{-15 }^{+20    }$ \\
    			Fe$^{+21}$ & $40    _{-13 }^{+20    }$ & $10    _{-10 }^{+30    }$ & $3.2   _{-3.2  }^{+9.4   }$ \\
    			Fe$^{+22}$ & $40    _{-20 }^{+60    }$ & $8.2   _{-8.2  }^{+30    }$ & $30    _{-20 }^{+30    }$ \\
    		\end{tabular}
    	\end{minipage}
    \end{table*}
    
\begin{figure*}
		\centering
		\includegraphics[trim=0 0 0.5cm 0,clip,width=\linewidth]{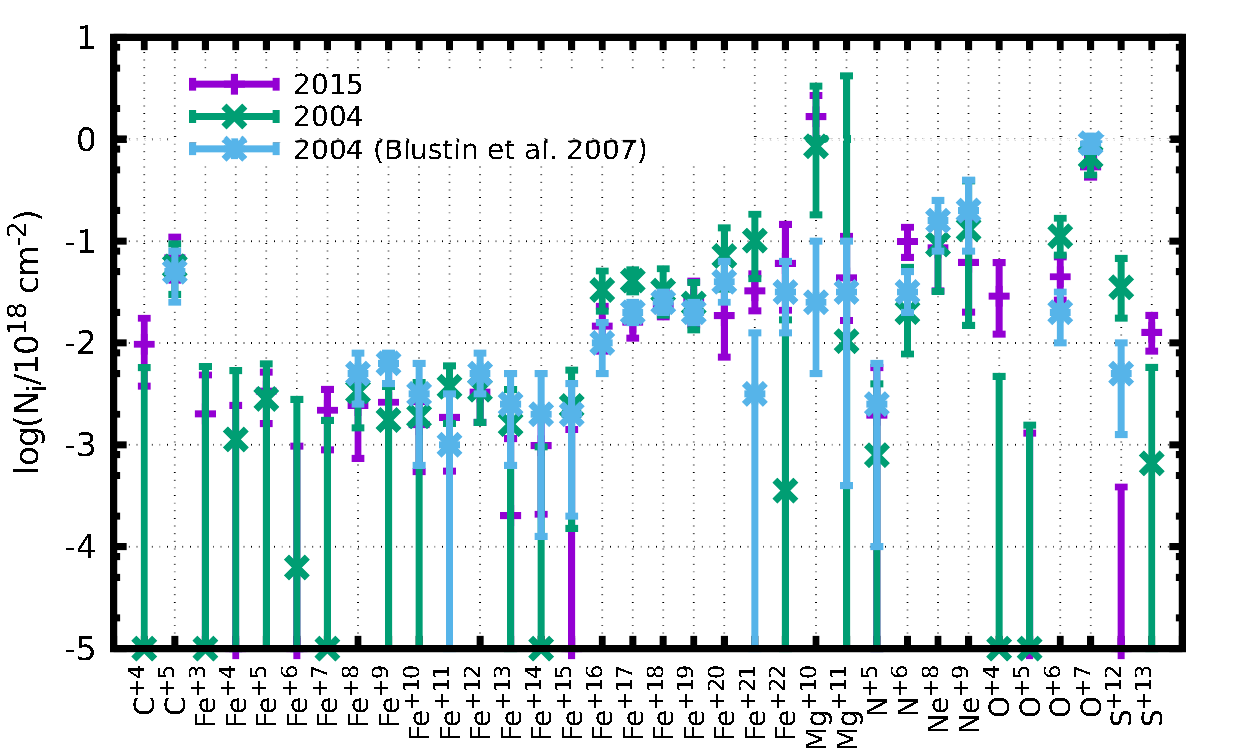}
		\caption{Comparison of column densities between 2004 and 2015 summed over the three velocity components
			along with the 90\% confidence intervals. The measurements by \citet{Blustin07} are also presented.
			Only N$^{+6}$,O$^{+4}$, Fe$^{+17}$, and S$^{+12}$ appear to have changed, but 3-4 ions are expected to deviate considering the 90\% uncertainties.}\label{fig:cols}
	\end{figure*}

    Finally the summed (across velocity components) column densities of the two epochs are given in Table \ref{tab:cols}.  
    These are compared graphically in Fig. \ref{fig:cols}, 
	as well as with the \citet{Blustin07} measured column densities for reference.
	While the different velocities may be associated with
	different physical components, 
	the current measurement is not sensitive to ionic column density changes in individual components due to the limited spectral resolution. 
	This is manifested in an inherit degeneracy of column densities between the velocity components, and the sum allows us to increase the sensitivity to change.
	
	A clear match can be seen, with 30/34 ion column densities within 90\% confidence.  
	Only N$^{+6}$, O$^{+4}$, Fe$^{+17}$, and S$^{+12}$ are discrepant between observations, but with 90\% uncertainties 3-4 measurements are expected to be discrepant. Moreover, other similar-ionization ions do not vary,
	indicating no absorption variability between the two epochs.
    
    \subsection{Absorption Measure Distribution}
    We characterize the ionization distribution of the absorber plasma using the Absorption Measure Distribution \citep[AMD][]{Holczer07}, defined as:
    \begin{equation}
    \mathrm{AMD}\triangleq\frac {\mathrm{d} N_\mathrm H} {\mathrm{d}\log \xi}
    \end{equation}
    where $N_\mathrm H$ is the column density and $\xi=L/(n_e R^2)$ is the ionization parameter. Here $n_e$ is the electron number density and $R$ is the distance of the absorber from the source.
    We can reconstruct the AMD using the measured ionic column densities:
    \begin{equation}
    N_\mathrm i=\int A_Z f_i(\xi) \frac {\mathrm{d} N_\mathrm H}{\mathrm{d}\log \xi}\mathrm{d}\log\xi\label{eq:niamd}
    \end{equation}
    where $A_Z$ is the solar abundance of the element \citep{Aspl09} and $f_i(\xi)$ is the fractional abundance of the ion as a function of $\xi$.
    We use a multiple thin shell model produced by XSTAR version
    2.38\footnote{\url{http://heasarc.gsfc.nasa.gov/docs/software/lheasoft/xstar/xstar.html}, along with AMD analysis code in
    \url{https://github.com/uperetz/AstroTools}, see README.} to determine the ionic fractions as a function of $\log \xi$. 
	The thin shell model assumes each $\log\xi$ is exposed to the unabsorbed continuum directly. This is justified by observing that the broad band continuum is not significantly attenuated by the absorption as seen by the relatively shallow edges (See Fig. \ref{fig:spectrum}).
	Our model grid is calculated from
    $\log \xi=-3.9$ to $\log \xi=3.9$ with $\Delta\log\xi=0.1$. 
    We use a Spectral Energy Distribution (SED) from 1 to 1000 Ry extrapolated from our multi-wavelength observations and corrected for galactic absorption 
    (M. Mehdipour et al., in preparation).
    
\begin{figure}
        \includegraphics[trim=0.5cm 0 1.5cm 1.1cm,clip,width=\linewidth]{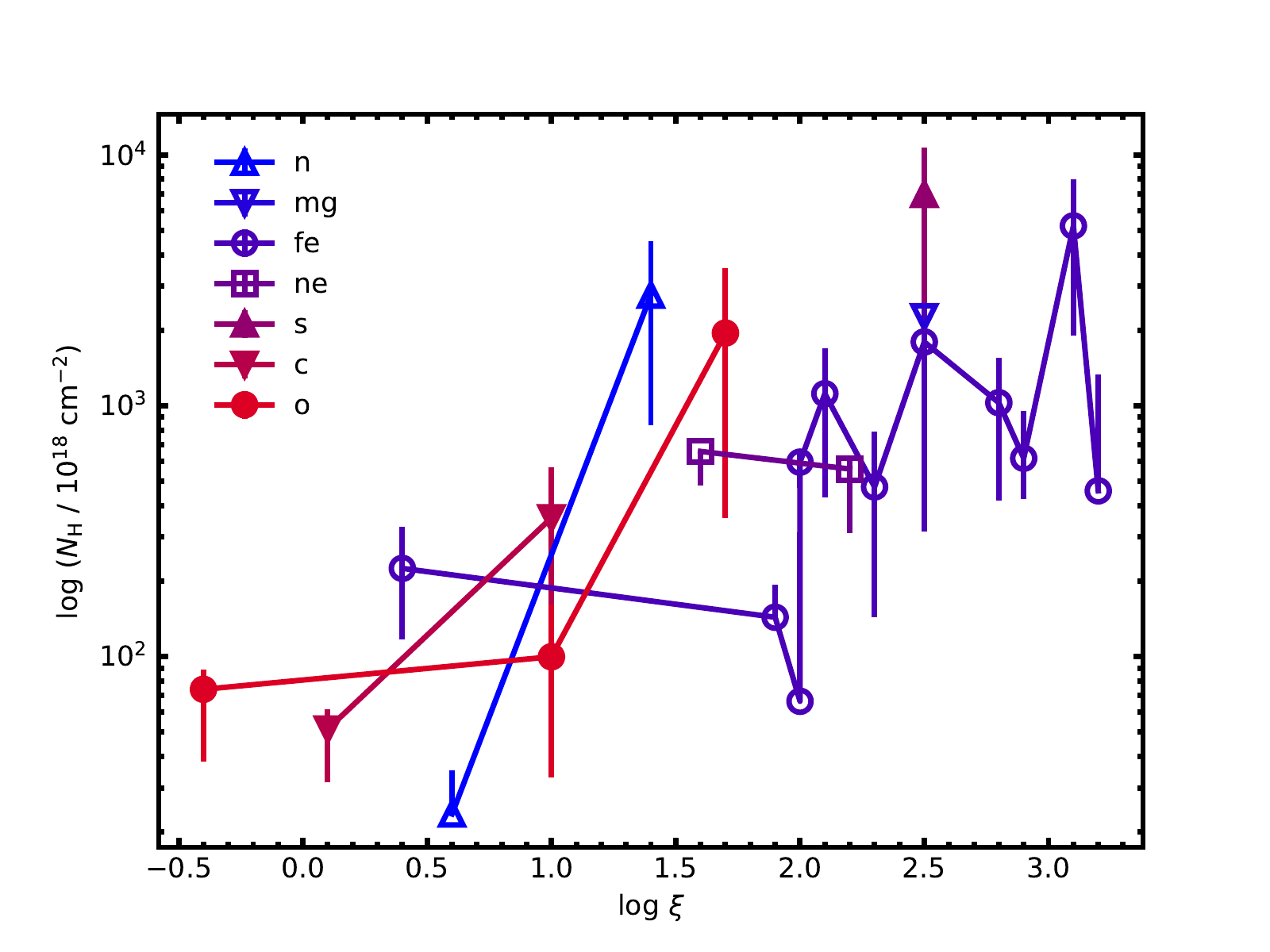}
        \caption{The distribution of $N_\mathrm H$ estimated from Eq. \ref{eq:ni}
        	for the --600 km/s velocity component of the outflow. 
        }\label{fig:amdest}
    \end{figure}

    
    An estimate of the AMD can be obtained assuming that each ion contributes its entire column
    at the $\xi_\mathrm{max}$ where the ion's relative ionic abundance peaks. 
    The total equivalent $N_\mathrm H$ for each $\log\xi_\mathrm{max}$ is then estimated by each ion:
    \begin{equation}
    N_\mathrm H = \frac{N_\mathrm i}{A_\mathrm Z f_i(\xi_\mathrm{max})}\label{eq:ni}
    \end{equation}
    This is a lower limit on column densities since in general $f(\xi)\leq f(\xi_\mathrm{max})$.
    The estimate is plotted in Fig. \ref{fig:amdest}, and shows a slight increase in column with $\log\xi$ consistent with \citet{Behar09}. Different ions from different elements in the same $\log\xi$ bin should agree, and discrepancies reflect deviations from solar abundances.
    
    In order to compute the AMD we want to solve the discretized set of equations (\ref{eq:niamd})
    \begin{equation}
    \textbf{N}=\widehat{A_Zf}~(\mathrm{\bf AMD}~\otimes\pmb{\Delta\log\xi})
    \end{equation}
    where $\widehat{A_Zf}$ is the matrix of ionic fractions given by XSTAR multiplied by $A_Z$, $\textbf{N}$  is the vector of measured ionic column densities, and $\mathrm{\bf AMD}~\otimes\pmb{\Delta\log\xi}$ is the vector of H column densities we want to find multiplied by the vector of AMD bins. Note the \textbf{AMD} vector is re-binned manually and may be uneven, enlarging the size of the bin until significant constraints are obtained for each bin. 
    The predicted columns are $\textbf{N}^\mathrm{p}=\widehat{A_Zf}~(\mathrm{\bf AMD}~\otimes\pmb{\Delta\log\xi})$. We use C-statistics \citep{Cash79} to fit  the AMD as we expect zero-value bins and there are less than 30 d.o.f.  We minimize the $C_\mathrm{stat}$ in order
    to find a best fit for the \textbf{AMD}:
    \begin{equation}
    C_\mathrm{stat}=2\sum_k^\mathrm{d.o.f.}(\textbf{N}^\mathrm p-\textbf{N}\ln\textbf{N}^\mathrm p)_k
    \end{equation}  
    The uncertainties of the measured ionic column densities are propagated stochastically. 
    We use 1000 Monte-Carlo runs on the vector $\textbf{N}$, where each column density is 
    rolled from a triangular probability distribution ranging through the 90\% confidence interval peaking at the best fit.
    
\begin{figure}[h!]
	\includegraphics[trim=0 0 4.5cm 2cm,clip,width=\linewidth]{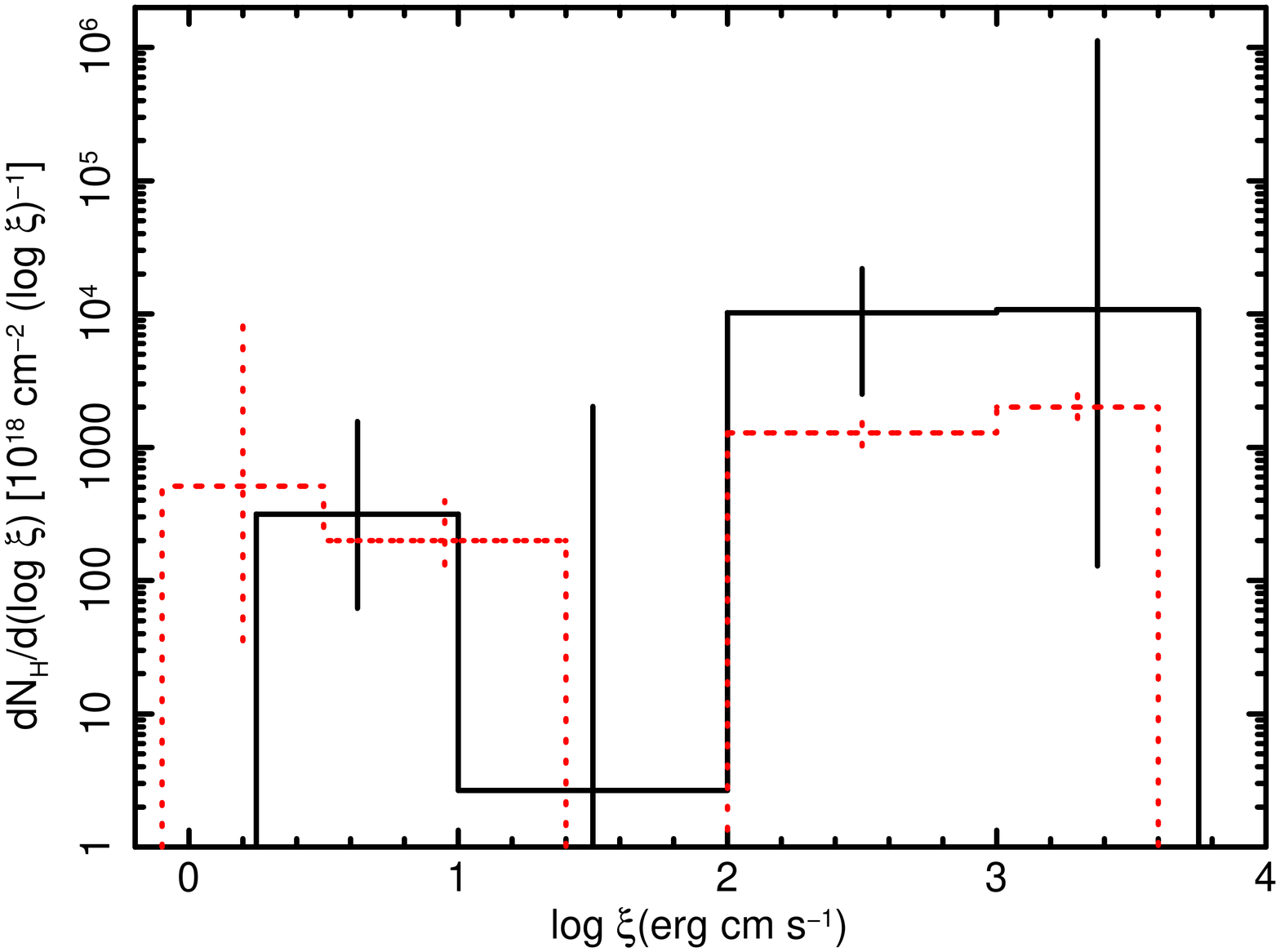}
	\caption{The AMD of NGC 7469 (solid) compared to that of \citet{Behar09} (dashed). The bins were determined such that a significant error may be obtained. The dip $\log\xi$ of 1 and 2 is attributed to thermal instability \citep{Holczer07}.}\label{fig:amd}
	\end{figure}

    The resulting AMD is plotted in Fig. \ref{fig:amd}, and resembles the AMD of \citet{Blustin07}.
    This is also well in agreement with the usual bi-modal shape commonly observed in AGNs \citep{Behar09,Laha14}.
    The consistency of the AMD structure along with the individual ionic column measurements increases our confidence that the absorber is unchanged between the 2004 and 2015 observations.
	
    \section{Variability and electron density}
    Following the previous work of \citet{Krolik95,Nicastro99}, and \citet{Arav12} we constrain a lower-limit on distance of the source to the outflow using the fact that no variability is measured in ionic column densities. From this we can estimate upper limits on $n_e$. In Appendix \ref{app:recomb} a rigorous derivation of the equations used in this section is provided for reference.

    \subsection{Days time scale variability}\label{sec:svar}
    The NGC~7469 lightcurve, created using the high statistics of the EPIC-pn, shows NGC~7469 has a variable continuum. In Fig. \ref{fig:pnlc} the 9  EPIC-pn lightcurves are presented, two from 2004 and the rest from 2015, with the count rate varying by up to a factor of 2 within a day. This rapid variability (compare with the year time scales, Sec. \ref{sec:lvar}) suggests the possibility of constraining the minimum response time to a change in ionizing flux of NGC~7469, and giving a lower limit on $n_e$ and thus an upper limit on the distance of the outflow from the AGN. This would only be possible if ionic column densities would be observed to change within the timescales of the continuum variability. In our case no variability can be detected on scales of days and longward, and thus  only lower limits on distance and upper limits on $n_e$ may be obtained.
    
    Since we can constrain the column densities at best to 50\%, evaluated by comparing the uncertainties to the best fit values, weaker variability is not ruled out.    
    Conversely, the lack of detected variability in $\sim30$ individual ions, as well as a lack of a systematic trend in the discrepancies between best-fit values, implies that if any change exists, it is small and may not be attributed to a change of the ionizing flux.   
    UV observations are more sensitive to variation in absorption troughs, and a detailed UV analysis of the epochs of NGC~7469 will be presented in a separate paper (Arav et al. in preparation).
    
\begin{figure*}
    \centering
	\includegraphics[width=\linewidth]{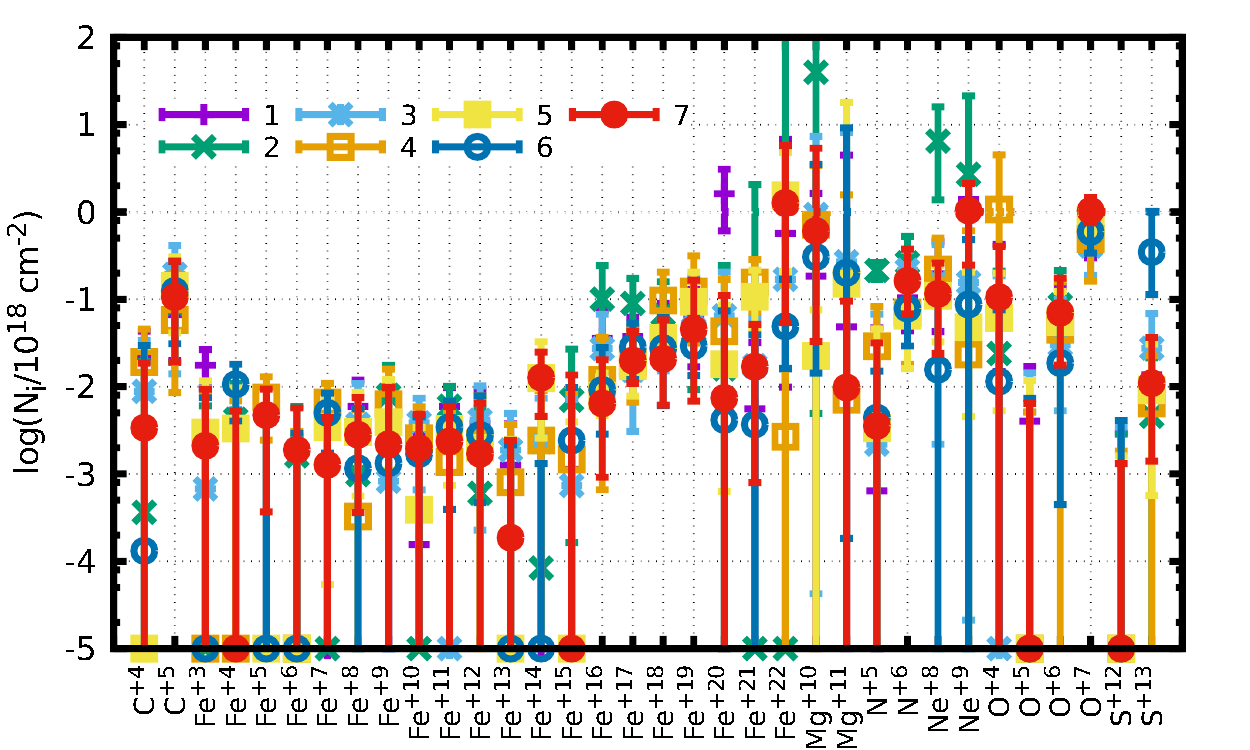}
	\caption{Comparison of the best-fit measured column densities for each individual epoch in 2015 (see Table \ref{tab:log}) summed over the three velocity components. While uncertainties may be large, measurements are consistent as being the same with no clear trend of change between observations. Variations of a single ion in a single epoch are unlikely to represent true variability.}\label{fig:scols}
	\end{figure*}

    In order to check the stability of the absorption due to the ionized outflow we apply the best-fit model on the combined spectra as a starting point for the fit of each individual spectrum. 
    Though the lower S/N of a single observation hampers tight constraints, the results are consistent within the 90\% uncertainty intervals across observations (Fig. \ref{fig:scols}), even better. The only exceptions are Ne$^{+8}$ and Fe$^{+20}$  deviating for one observation, but not the same one. Beyond constancy among observations, when considering the best-fit values it is evident that there is no clear trend - the ordering of column densities of different ions of similar ionization parameter between observations is not uniform. This indicates there is no observable change, in fact, of the ionic column densities during the last half year of 2015.
    
    \subsection{Intra-day time scale variability: comparing high and low states }
    \begin{figure}
        \centering
        \includegraphics[trim=0.7cm 0.35cm 5cm 2.5cm,clip,width=0.95\linewidth]{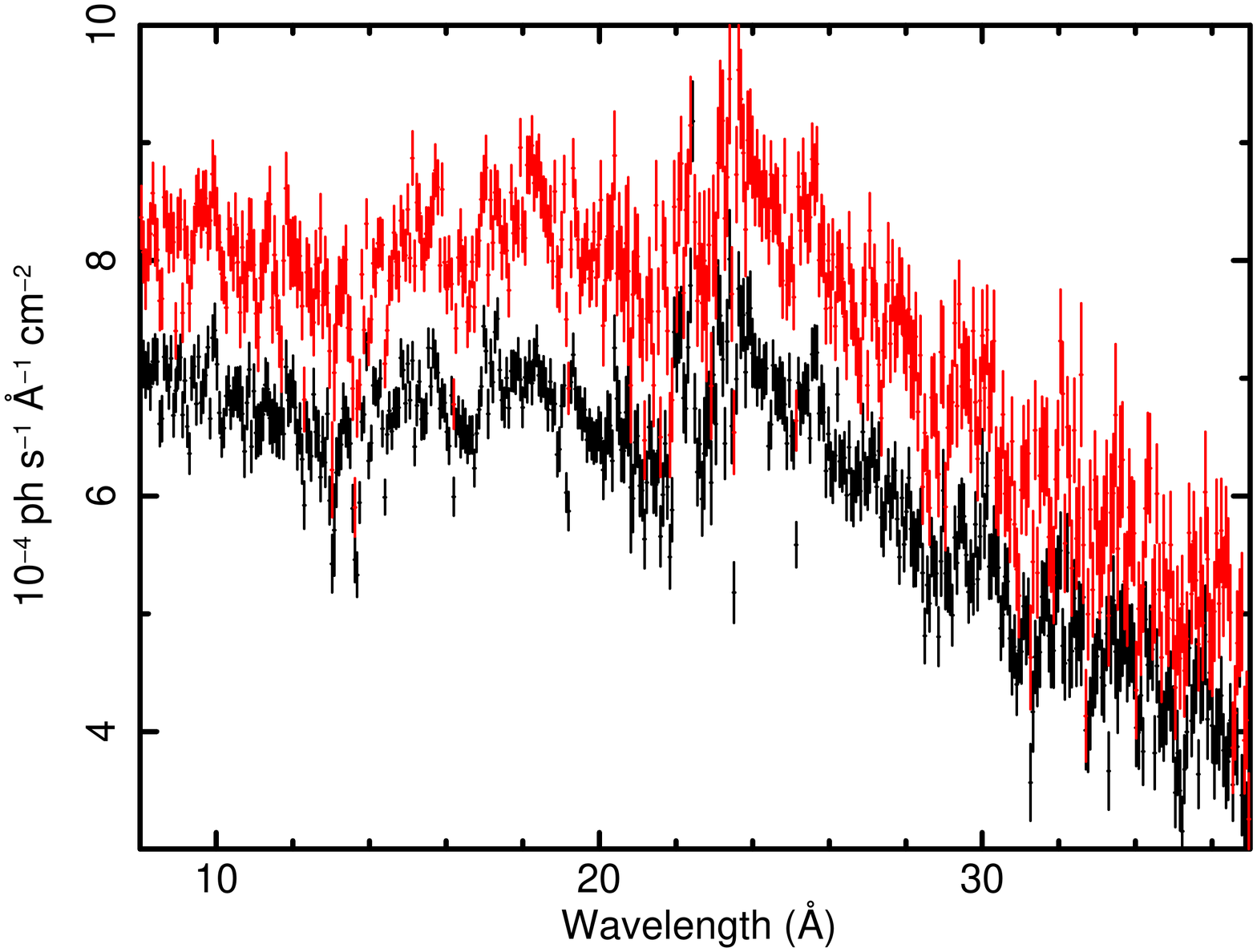}
        \caption{High and low state spectra of NGC 7469 during 2015. Spectra have been binned to 60 m\AA. Bottom panel shows residuals to the best-fit galactically-absorbed powerlaw. No obvious change in lines is observed, see Fig. \ref{fig:statecmp} for a detailed comparison of ionic column densities.}
        \label{fig:states}
    \end{figure}
    The stochastic nature of the ionizing flux may lead to an hypothesis that any outflow which is not dense and close to the source would not respond quickly enough to the changes, at least not measurably. 
    By summing spectra of predominantly high and low states of the AGN separately,
    more subtle changes can be measured 
    by improving the S/N of small absorption troughs which change in a consistent manner, on a daily basis.
    
\begin{figure*}
        \centering
        \includegraphics[width=\linewidth]{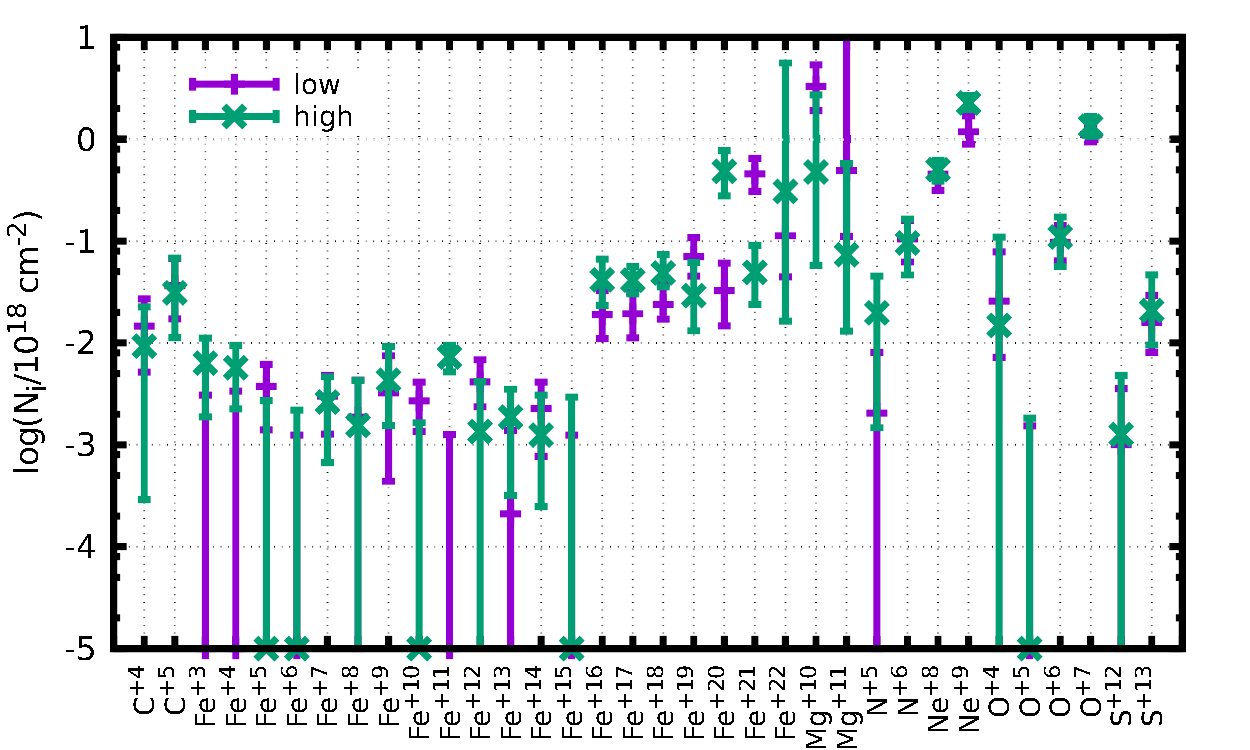}
        \caption{Ionic column density measurements of the low and high states of NGC 7469 during 2015 summed over the three velocity components. 
            Fe$^{+20}$ and Fe$^{+21}$ interestingly enough seem to exhibit mutual variability, but it is opposite to what is expected - higher state leading to higher ionization, as well as no other ion displaying such behavior. With 90\% confidence 3-4 ions are expected to deviate.}\label{fig:statecmp}
    \end{figure*}

    We divide the states according to the EPIC-pn lightcurve, around the mean count rate for the 2015 observations
    (which is nearly identical to the median one) of 23.2 counts s$^{-1}$. 
    Retaining all photons in favor of statistics and in order to secure similar RGS S/N in the high and low states, we cut the events 
    at $0.4\sigma$ above the mean EPIC-pn count rate ($\sigma=3.5$ counts s$^{-1}$ is the 
    standard deviation of the light curve). 
    These spectra are presented in Fig. \ref{fig:states}, showing very similar troughs. As was the case for the individual epoch analysis, we begin a fit from the best-fit model of the combined spectra. Results are presented in Fig. \ref{fig:statecmp}. Here 3 ions only deviate, which is expected within the 90\% statistics.
    
    Once again the NGC 7469 outflow proves to be remarkably stable such that when observing only times of high flux and comparing to times of low flux, no change is observed in column densities and thus the outflow ionization distribution. Here, variability is constrained at best to 25\% (by comparing the uncertainties to the best fit values), and again, weaker variations may be present.
    
    \subsection{Year time scale variability}\label{sec:lvar}
    While on timescales of days and less we see that the continuum of NGC 7469 is variable in EPIC-pn lightcurves (Fig. \ref{fig:pnlc}), ionic column densities remain unchanged over timescales
    of days and months, observed during 2015. In addition, the column densities are comparable to those of 2004, despite the 25\% difference in flux (See Table \ref{table:continuum}). 
    Thus, we make the assumption that column densities remain unchanged for the entire $T=10$ years.
    This assumption allows us to constrain the distance $R$ of the outflow from the ionizing
    source assuming $\tau>T$, where $\tau$ is the ionization equilibrium time \citep{Krolik95,Nicastro99,Arav12}. 
    
    A full derivation of the dependence of $R$ on $T$ is detailed in Appendix \ref{app:recomb}.
    While the power of this derivation cannot be fully utilized for NGC~7469 as we detect no variability, a useful result for this case is:
    \begin{equation}\label{eq:rmaxpaper}
        R^2>\left(\frac {\alpha_i L} {\xi_\mathrm{max}}+\mathcal{L}_i\right)T      
    \end{equation}
    Given an ion $i$ the recombination rate coefficient is $\alpha_i$. 
    The photoionization cross-section and rate are, respectively, $\sigma_i$ and $\mathcal L_i/R^2$, where
    \begin{equation}
    \mathcal{L}_i=\frac 1 {4\pi} \int_{1 \mathrm{Ry}}^{1000 \mathrm{Ry}}\frac{\sigma_i(E)L(E)}{E}\mathrm{d} E 
    \end{equation}
	$L(E)$ is the luminosity density in erg s$^{-1}$ keV$^{-1}$ and $L$ is the ionizing AGN luminosity:
	\begin{equation}
	L=\int_{1\mathrm{Ry}}^{1000\mathrm{Ry}}L(E)\mathrm{d} E
	\end{equation}	
    $L(E)$ and $L$ are estimated from the SED, which yields $L=1.4\times 10^{44}$ erg s$^{-1}$. Recombination and photoionization coefficients are taken from the CHIANTI software package\footnote{http://www.chiantidatabase.org/} \citep{chianti}.
    Once we obtain a lower limit on the distance, we may use the definition of $\xi$ to extract an upper limit on $n_e$:
    \begin{align}
    n_e < \frac L {\xi R^2_{\mathrm{min}}}
    \end{align}
    where $R_{\mathrm{min}}$ is the minimal value obtained from eq. \ref{eq:rmaxpaper}, and the same $\xi$ and $L$ values are used.
    
    Distances and electron densities measured from several ions are given in Table \ref{table:distance}. The outflow is constrained to be at least 12 pc away from the source for Fe$^{+22}$ and  31 pc for N$^{+6}$. This constraint is not strong enough to dis-associate it from the AGN completely, or associate it with the starburst region seen in NGC 7469 \citep{David92}, which is approximately 1 kpc from the source.
    
    \begin{table*}
    	\centering
    	\caption{Calculated parameters and limits on outflow properties}\label{table:distance}
    	\begin{tabular}{c|c|c|c|c|c|c}
    		ion & $\log \xi$        & $\alpha_i$                 & $\alpha_i L/\xi_i$        & $\mathcal{L}_i$                    & $R>$  & $n_e<$\\
    		& erg cm s$^{-1}$ & $10^{-12}$ cm$^3$ s$^{-1}$ & $10^{30}$ cm$^2$ s$^{-1}$ & $10^{30}$ cm$^2$ s$^{-1}$ & pc    & cm$^{-3}$\\
    		\hline
    		&&&&&&\\[-0.25cm]
    		N$^{+6}$  &$1.4^{+1.1}_{-0.9}$& $1.5^{+0.4}_{-0.5}$ & $8.1^{+45}_{-7}$  & $21.7$& $31^{+18}_{-4}$& $590^{+3280}_{-540}$\\
    		O$^{+7}$  &$1.7^{+0.7}_{-0.6}$& $1.6^{+0.3}_{-0.4}$ & $4.4^{+16}_{-3.7}$& $9.8$ & $22^{+10}_{-3}$& $620^{+2750}_{-570}$\\
    		Fe$^{+22}$&$3.2^{+0.1}_{-0.2}$& $35^{+64}_{-19}$    & $3.1^{+14}_{-1.9}$& $1.7$ & $12^{+13}_{-3}$& $59^{+140}_{-48}$\\
    	\end{tabular}    
    \end{table*}
    
    \section{Energy deposit}
    Using mass conservation in a continuous conical outflow with opening angle $\Omega$, $\mathrm{d} m=\Omega R^2 n_e v \mu m_p \mathrm{d} t$, 
    we define the kinetic power of the outflow as
    \begin{equation}\label{eq:ekdot}
    \dot{E}_K=\frac 1 2 \frac{\mathrm{d} m}{\mathrm{d} t} v^2=\frac{\Omega}{2} n_e R^2 v^3 \mu m_p=\frac{\Omega}{2} \frac L \xi v^3 \mu m_p
    \end{equation}
    Where $\mu=1.4$ is the mean molecular weight and $m_p$ is the proton mass. 
    We assuming here a bi-conical flow of $\Omega=2\pi$.
    We use the maximal velocity component of $-2000$ km s$^{-1}$   
    and the lowest ionization observed at that velocity, $\log\xi=1$ (excluding ions with column density consistent with 0). These values are seen for example in C$^{+5}$ and O$^{+6}$. This yields a maximal possible value of
    \begin{equation}\label{eq:superhigh}
    \dot{E}_K=8.2\times10^{44} \mathrm{\ erg\ s}^{-1}\approx0.6L_{Edd}
    \end{equation}
    where $L_\mathrm{Edd}=1.4\times10^{45}$ erg s$^{-1}$.
    Using a high $\log\xi=2.5$ but leaving the velocity of  $-2000$ km s$^{-1}$ (observed for example in Fe$^{+22}$) will reduce this value by two orders of magnitude:
	\begin{equation}\label{eq:superlow}
	\dot{E}_K=6.2\times10^{42} \mathrm{\ erg\ s}^{-1}\approx0.004L_{Edd}
	\end{equation}
    Substituting in the lowest velocity of $-600$ km s$^{-1}$ will reduce $\dot E_K$ by another 1.5 orders of magnitude, and an opening angle less than $2\pi$ would reduce it even further.
    
    The fact that a range of $\xi$ values is ubiquitously observed in AGN outflows
    indicates the wind cannot have a conical $n_e\propto R^{-2}$ density profile.
    Multiple ionization winds have been discussed in the models of \citet{Fukumura10,Stern14}.
    Eq. \ref{eq:ekdot} results in an increase of power with decreasing ionization.
    
    Other definitions of kinetic luminosity, such as that of \citet{Borguet12}, assume a thin shell of thickness $\Delta R<R$ rather than a continuous outflow, dividing the mass by the traversal timescale, $R/v$. In that case the kinetic luminosity would be lower by $\Delta R/R$.
    
    One may also assume $\Delta R=R$, such that $\dot E_K\propto n_eR^2=N_HR$.
    In this case we can use the measured lower limits on distance (Table \ref{table:distance}) and the measured equivalent H column densities (Eq. \ref{eq:ni}).
    Lower bounds for $\dot E_K$ from N$^{+6}$, O$^{+7}$, Fe$^{+22}$ respectively are
    $1.1\times10^{43},~4.2\times10^{41}$, and $2.6\times10^{42}$  erg s$^{-1}$. 
    Note for each ion we use the fastest velocity where measured column density is inconsistent with 0, namely --600 km s$^{-1}$ for N$^{+6}$, and --2000 km
    s$^{-1}$ for O$^{+7}$ and Fe$^{+22}$. The lowest estimate is even lower than that of Equation \ref{eq:superlow}.
    
    Assuming the highest estimate of the kinetic power (Eq. \ref{eq:superhigh}) is the true energy carried by the outflow would imply significant feedback.
    However, the fact is that a starburst region 
    is observed at 1 kpc \citep{David92} and does not seem to be affected by the 
    outflow. This would lead to the conclusion that the outflow is spatially de-coupled
    from the starburst region. 
    If the outflow power is much lower as in Eq. \ref{eq:superlow}, this 
    would naturally explain why the starburst region is unaffected.

    \section{Conclusions}
    The X-rays absorption spectra of NGC~7469 is remarkably stable on all of the measured time scales. In observations spread over years, months and days column densities associated with the ionized absorber are not observed to change. 
    On the other hand, the intrinsic variability of the source is large, changing by up to a factor of two in the course of a single day.
    In addition, the average soft X-ray powerlaw slope changes between 2004 and 2015 from 2.1 to 2.3, again, with no observed absorption variability.
    
    The kinematic components of the outflow are also constant between the 2004 and 2015 observations, and between the X-ray and the UV bands.    
    Constancy of the outflow can also be observed in the reconstructed AMD, featuring one high ionization component and one low ionization component with the same column densities in both 2004 and 2015. Admittedly, the broad and relatively flat AMD makes ionization changes much harder to detect than in a single-$\xi$ component. To that end, we would expect to notice changes only in the highest and lowest ionization states. Nonetheless, the UV spectra of this campaign (Arav et al. in preparation) confirms for the most part the lack of absorption variability, except for minor changes that are detected in a few velocity bins in the UV, but are much below the current X-ray sensitivity.
    
    The flux variations on different timescales with no effect whatsoever on the outflow imply a distant outflow, several pc away from the AGN at least. Beyond the large distance, the velocities, luminosity, and observed ionization parameters suggest
    the outflow may carry as much as 2/3 of the Eddington AGN power, which is significant in terms of feedback.
    However, this is dependent on $\xi$ (Eq. \ref{eq:ekdot}) as expected for non-conical outflows, and is 2 orders of magnitude lower for high $\xi$ values, making these estimates
    ambiguous and inconclusive as estimators of feedback without a physical model associated with $\dot E_K$ .
   
    We found no evidence the AGN is responsible for driving the outflow, since the distance scales are beyond the torus \citep{Suganuma06} and comparable to the region of narrow ($\sim500$ km s$^{-1}$) line emission. The obtained constraints on distance and power of the outflow need to be examined in other AGNs in order to understand if these outflows are unimportant to the galactic scale, and what is their connection to the AGN itself.
    
    \begin{acknowledgements}
    This work was supported by NASA grant NNX16AC\-07G through the {\it XMM-Newton} Guest Observing Program, and through grants for HST program number 14054 from the Space Telescope Science Institute, which is operated by the Association of Universities for Research in Astronomy, Incorporated, under NASA contract NAS5-26555. 
    The research at the Technion is supported by the I-CORE program of the Planning and Budgeting Committee (grant number 1937/12). 
    EB received funding from the European Union's Horizon 2020 research and innovation programme under the Marie Sklodowska-Curie grant agreement no. 655324. 
    SRON is supported financially by NWO, the Netherlands Organization for Scientific Research.
    NA is grateful for a visiting-professor fellowship at the Technion, granted by the Lady Davis Trust.   
    SB and MC acknowledge financial support from the Italian Space Agency under grant ASI-INAF I/037/12/0.
    BDM acknowledges support from the European Union's Horizon 2020 research and innovation programme under the Marie Sk\l odowska-Curie grant agreement No. 665778 via the Polish National Science Center grant Polonez UMO-2016/21/P/ST9/04025.
    LDG acknoweledges support from the Swiss National Science Foundation.
    GP acknowledges support by the Bundesministerium f\"{u}r Wirtschaft und 
    Technologie/Deutsches Zentrum f\"{u}r Luft- und Raumfahrt 
    (BMWI/DLR, FKZ 50 OR 1408 and FKZ 50 OR 1604) and the Max Planck Society.
    POP acknowledges support from CNES and from PNHE of CNRS/INSU.
    \end{acknowledgements}

    \begin{appendix}
    \section{Equilibrium time}\label{app:recomb}
    Following the works of \citet{Krolik95,Nicastro99,Arav12} we define the two inverse timescales for ionization and recombination respectively:
    \begin{align}
    \nonumber
    \mathcal{J}_i&=\int_{\nu_0}^{\infty}\frac{\sigma_i(\nu)J(\nu)}{h\nu}\mathrm{d}\nu= \\ 
        &=\frac 1 {4\pi R^2}
    \int_{\nu_0}^{\infty}\frac{\sigma_i(\nu)L(\nu)}{h\nu}\mathrm{d}\nu \triangleq \frac {\mathcal{L}_i} {R^2}\\
    \mathcal{R}_i&=\alpha_i n_e
    \end{align}
    The ionization/recombination/equilibrium time $\tau$ used in this paper is the decay time of the exponential solution of the system of equations for the ionic populations $n_i$:
    \begin{equation}
    \dot{n}_i=-(\mathcal{J}_i+\mathcal{R}_i)n_i+\mathcal{J}_{i-1} n_{i-1}+\mathcal{R}_{i+1} n_{i+1}\label{eq:nidot}
    \end{equation}
    for charge states $0<i<Q$ and the boundary defined by $n_{Q+1}=n_{-1}=0$ or: 
    \begin{align}
    \dot n_0=&-\mathcal{J}_0n_0+\mathcal{R}_1 n_1\label{eq:dotn0}\\
    \dot n_Q=&-\mathcal{R}_Qn_Q+\mathcal{J}_{Q-1}n_{Q-1}
    \end{align}
    These equations assume all charge states are exposed to the same radiation field $J(\nu)$.
    In the general case where the radiation field $J(\nu)$ is non-uniform this approximation breaks down.
    
    \subsection{Assumptions and caveats}\label{sec:assum}
    In general, \ref{eq:nidot} must be solved for a  time varying set of $\mathcal{J}_i,\mathcal{R}_i$, making the full solution much more difficult, and is formally given in \citet{Krolik95}. This is less practical when we want to use our measurements to constrain unobserved quantities, such as $n_e$. In this case, we often want to consider a system in equilibrium, with a given inital set $\mathcal{J}_{i}^0,\mathcal{R}_{i}^0$, where we abruptly change the external conditions using a new set of $\mathcal{J}_{i}^{\mathrm {final}}$ - making the assumption that the continuum changed as a step function, and we are observing much after the step (See Sec. \ref{sec:necalc}), or conversely that the
    system is in equilibrium and this abrupt change has yet to be observed. Though often not the case, this is a good assumption when observing the outflow much before and much after such a change in seed flux, such that the continuum observed is steady for times greater than the $\tau$. Consider now a short scale oscillating variation in seed flux,
    \begin{equation}
    t_{short}\ll \tau
    \end{equation} 
    AGNs in general (indeed, NGC 7469 is a good example) may change drastically on timescales of days, with no observable change in column densities. In this case we may assume that the effective continuum on the plasma is in fact a steady one, given by the time averaged flux.
    \begin{align}
    \nonumber
        \overline{\mathcal{J}}_i =& \frac{\int_0^{T\ge t_{short}}\mathcal{J}_i \mathrm{d} t}{T\ge t_{short}}=\frac 1 T \int_0^T \mathrm{d} t
        \int_{\nu_0}^{\infty}\frac{\sigma_i(\nu)J(\nu)}{h\nu}\mathrm{d}\nu=\\
        =&\int_{\nu_0}^{\infty}\frac{\sigma_i(\nu)\overline J(\nu)}{h\nu}\mathrm{d}\nu
    \end{align}
    So we make 3 assumptions when analyzing this photoionized plasma:
    \begin{enumerate}
        \item If no column densities are observed to change while flux varies on short (hours,days) time scales, a steady time averaged continuum may be assumed.
        \item If column densities are changed between two observations, and the flux is shown to be steady, we will assume $T_{final}>T_{start}+\tau$ where $T_{final}$ is the final observation and $T_{start}$ is the time where the continuum started to change, after the first observation. In this case we assume a step function change for the continuum.
        \item Finally, if column densities remain unchanged between two observations but flux is shown to have changed and remain steady, we will assume $\tau>\delta T$, the time between observations.
    \end{enumerate}
    \subsection{Solution}\label{sec:solution}
    From the form of the equations, or from solving the simple 2-level system one may quickly come to the conclusion a general solution should be of the form (assuming constant $\mathcal{J}_i,\mathcal{R}_i$, as per Sec. \ref{sec:assum}):
    \begin{equation}\label{eq:form}
    n_i=A_i\hspace{0.05cm}\mathrm{\textbf{e}}^{-\frac t \tau}+B_i
    \end{equation}
    First-order differential equations have only one free coefficient depending on the initial conditions. 
    $\tau$ must be independent of charge, and this can easily be shown by substituting different $\tau_i,\tau_j$ for consecutive charge states into the equation for $\dot n_i$, \ref{eq:nidot}, assuming $A_i$ and $B_i$ are constants. 
    Some properties of this solution are evident immediately. Assuming steady state before $t=0$ and at $t\rightarrow\infty$ leads to the conclusion:
    \begin{align}
    B_i&=n_{fi}\\
    A_i&=n_{ii}-n_{fi}
    \end{align}
    where $n_{fi}$ are the equilibrium densities at $\infty$ and $n_{ii}$ are the initial equilibrium densities. An important consequence is that $B_i$ are not integration coefficients. These are the final equilibrium solutions, explicitly given by $\mathcal{J}_i,\mathcal{R}_i$, as seen in the Section \ref{app:equi}. Substitute in our form \ref{eq:form} to \ref{eq:nidot}:
    \begin{align*}
    -\tau^{-1}A_i\hspace{0.05cm}\mathrm{\textbf{e}}^{-\frac t \tau}=&-(\mathcal{J}_i+\mathcal{R}_i)(A_i\hspace{0.05cm}\mathrm{\textbf{e}}^{-\frac t \tau}+B_i)+ \\
    &+\mathcal{J}_{i-1}(A_{i-1}\hspace{0.05cm}\mathrm{\textbf{e}}^{-\frac t \tau}+B_{i-1})+\\
    &+\mathcal{R}_{i+1}(A_{i+1}\hspace{0.05cm}\mathrm{\textbf{e}}^{\frac t \tau}+B_{i+1})\rightarrow\\
    \end{align*}
    \vspace{-1.2cm}
    \begin{align*}
    \rightarrow A_{i+1}\hspace{0.05cm}\mathrm{\textbf{e}}^{\frac t \tau}&+B_{i+1}=\\
    &=\frac 1 {R_{i+1}}\left(\left(\mathcal{J}_i+\mathcal{R}_i-\tau^{-1}\right)A_i-\mathcal{J}_{i-1}A_{i-1}\right)\hspace{0.05cm}\mathrm{\textbf{e}}^{-\frac t \tau}+\\
    &+\frac 1 {R_{i+1}}\left(\left(\mathcal{J}_i+\mathcal{R}_i\right)B_i-\mathcal{J}_{i-1}B_{i-1}\right)
    \end{align*}
    Grouping the coefficient for the exponent and constant results in the formulas for the coefficients:
    \begin{align}
    A_{i+1}&=\frac 1 {R_{i+1}}\left(\left(\mathcal{J}_i+\mathcal{R}_i-\tau^{-1}\right)A_i-\mathcal{J}_{i-1}A_{i-1}\right)\label{eq:A}\\
    B_{i+1}&=\frac 1 {R_{i+1}}\left(\left(\mathcal{J}_i+\mathcal{R}_i\right)B_i-\mathcal{J}_{i-1}B_{i-1}\right)\label{eq:B}  
    \end{align}
    It is easy to prove that \ref{eq:B} results in $B_i/B_{i-1}=\mathcal{J}_{i-1}/\mathcal{R}_i$ which we know must be true, as $B_i$ are an equilibrium solution (see Sec. \ref{app:equi}). What will be interesting to us is the relation of $A_i$ to $\tau$.
    \subsection{Equilibrium time}\label{sec:necalc}
    A closed form solution for $A_i$ is more difficult, but we are only interested in $\tau$, which may be obtained from \ref{eq:A} using any observed ionization triad:
    \begin{equation}
    \begin{split}
    \tau=&\left((\mathcal{J}_i+\mathcal{R}_i)-\frac{\mathcal{R}_{i+1}A_{i+1}+\mathcal{J}_{i-1}A_{i-1}}{A_i}\right)^{-1}=\\
    =&\left(\vphantom{\frac L \xi} (\mathcal{J}_i+\mathcal{R}_i)- \right. \\
    -&\left. \frac{\mathcal{R}_{i+1}(n_{ii+1}-n_{fi+1})+\mathcal{J}_{i-1}(n_{ii-1}-n_{fi-1})}{n_{ii}-n_{fi}}\right)^{-1}
    \end{split}\label{eq:trmid}
    \end{equation}
    Measuring 3 ions of an element and seed flux of 2 different observation epochs will allow us to constrain $n_e$. In terms of what we measure:
    \begin{equation}
    \frac{n_{ii-1}-n_{fi-1}}{n_{ii}-n_{fi}}=\frac{\delta n_{i-1}}{\delta n_i}=l\frac{\delta N_{i-1}}{\delta N_i}
    \end{equation}
    where $N_i$ are the column densities of the specific ions and $l$ is the ratio of widths over which the two ions extend. We will assume $l=1$ as $\xi$ is inversely proportional to $R$ and \citet{Kallman96} shows most adjacent ion stages tend to extend over similar $\xi$ ranges, and indeed may exist in the same part of the plasma, though this does not have to be the case. 
    
    An interesting thing to note is that the equilibrium constants $n_{i,f}$ are also dependent on the $\mathcal{R}$ and $\mathcal{J}$, and obviously each is a different set of constants as both $n_e$ and $J(\nu)$ have changed, but only those of $n_{fi}$ are the same as the explicit $\mathcal{J}$ and $\mathcal{R}$ appearing in \ref{eq:trmid}. Finally, substituting the expressions for $\mathcal{J}_i,\mathcal{R}_i$ obtain the relationship we need:
    
    \begin{equation}\label{eq:tausol}
    \begin{split}
    \tau=&\left(\left(\alpha_i-\frac{\delta N_{i+1}}{\delta N_i}\alpha_{i+1}\right)n_e+ \right. \\
    &+\left. \left(\mathcal{L}_i-\frac{\delta N_{i-1}}{\delta N_i}\mathcal{L}_{i-1}\right)R^{-2}\right)^{-1}
    \end{split}
    \end{equation}
    We note this equation is the same as eq. 10 in \citet{Arav12} when 
    \begin{equation}
    \begin{split}
    &\left(\mathcal{L}_i-\frac{\delta N_{i-1}}{\delta N_i}\mathcal{L}_{i-1}\right)R^{-2}=\\
    =&-\frac{J(t>0)}{J(t=0)}\left(\alpha_i-\frac{\delta N_{i+1}}{\delta N_i}\alpha_{i+1}\right)n_e
    \end{split}
    \end{equation}
    and $\delta N = N$, tying a step change in ionization flux to recombination.
    
    Note that the ionization parameter is an observable that is found independently:
    \begin{equation}\label{eq:xi}
    \xi=\frac L {n_e R^2}= \frac {\int_{1 \mathrm{Ry}}^{1000 \mathrm{Ry}} L(\nu)\mathrm{d}\nu} {n_e R^2}
    \end{equation}
    While at first glance this may seem like it would be embedded somehow in \ref{eq:tausol}, note that $\xi$ is a purely equilibrium characteristic of the plasma, while $\tau$ is of course the time scale characterizing the system out of equilibrium. This gives us physical justification to say $\ref{eq:tausol}$ and $\ref{eq:xi}$ are independent equations, and may be solved simultaneously for $n_e$ and $R^2$:
    \begin{equation}
       \begin{split}
        n_e=&\left(\left(\alpha_i-\frac{\delta N_{i+1}}{\delta N_i}\alpha_{i+1}\right)+ \right. \\        
        &+\left. \left(\mathcal{L}_i-\frac{\delta N_{i-1}}{\delta N_i}\mathcal{L}_{i-1}\right)\frac \xi L\right)^{-1}\tau^{-1}
        \end{split}
    \end{equation}
    \vspace{-0.5cm}
    \begin{equation}
        \begin{split}    
        R^2=&\left(\left(\alpha_i-\frac{\delta N_{i+1}}{\delta N_i}\alpha_{i+1}\right)\frac L \xi+ \right. \\
        &+\left. \left(\mathcal{L}_i-\frac{\delta N_{i-1}}{\delta N_i}\mathcal{L}_{i-1}\right)\right)\tau
        \end{split}
    \end{equation}
    An interesting consequence is that the coefficients of $\tau$ must be positive. If this is not the case, then these solutions are wrong and our assumptions need to be put to test. Note that for a two level system this must be true as the ratio of column change is always negative.
    
    \subsection{Applications}
    While most parameters insofar are either measurable independently ($\mathcal{L}_i,L,\delta N_i,\xi$) or known ($\alpha_i$) we in general only have a limit on $\tau$ as we do not observe the plasma continuously.
    To practically apply this result to observational data we need inequalities, not equalities. Assume we know $\tau$ is lower than some constant $T$, a time between two observations. This happens often when we see an AGN in a steady low/high state at one time, and a high/low in another, with different columns. We can then use:
    \begin{equation}
        \begin{split}
        n_e>&\left(\left(\alpha_i-\frac{\delta N_{i+1}}{\delta N_i}\alpha_{i+1}\right)+\right. \\
        &+\left. \left(\mathcal{L}_i-\frac{\delta N_{i-1}}{\delta N_i}\mathcal{L}_{i-1}\right)\frac \xi L\right)^{-1}T^{-1}\\
        \end{split}
    \end{equation}
    \vspace{-0.5cm}
    \begin{equation}
        \begin{split} 
        R^2<&\left(\left(\alpha_i-\frac{\delta N_{i+1}}{\delta N_i}\alpha_{i+1}\right)\frac L \xi+ \right. \\
        &+\left. \left(\mathcal{L}_i-\frac{\delta N_{i-1}}{\delta N_i}\mathcal{L}_{i-1}\right)\right)T
        \end{split}   
    \end{equation}
    
    to constrain a maximal $R$, and minimal electron density. If on the other hand no variability is measured we are struck with a problem. While we would know $\tau>T$, so constraints would be reversed, we do not know the final column densities. One way to handle this is to make the assumption ${\delta N_{i+1}}/{\delta N_i}$ is, as a two level system, always negative, allowing an estimate:
    \begin{align}\label{eq:rmax}
        \nonumber
        R^2>&\left(\left(\alpha_i-\frac{\delta N_{i+1}}{\delta N_i}\alpha_{i+1}\right)\frac L \xi+\left(\mathcal{L}_i-\frac{\delta N_{i-1}}{\delta N_i}\mathcal{L}_{i-1}\right)\right)T>\\
        >&\left(\frac {\alpha_i L} \xi+\mathcal{L}_i\right)T
    \end{align}
	and consequently following from eq. \ref{eq:xi} we have:
	\begin{align}
	n_e < \frac L {\xi R^2_{\mathrm{min}}}
	\end{align}
	where $R_{\mathrm{min}}$ is obtained from the lower limit given by eq. \ref{eq:rmax}. This is the approximation used in Sec. \ref{sec:lvar}.
	
    \subsection{Equilibrium}\label{app:equi}
    We add this section for completeness' sake only. This problem can trivially be solved for the case $\dot n_i=0$, where by induction if ${n_{i-1}}/{n_i}={\mathcal{R}_{i}}/{\mathcal{J}_{i-1}}$ and
    \begin{equation}
    0=-(\mathcal{J}_i+\mathcal{R}_i)n_i+\mathcal{J}_{i-1}n_{i-1}+\mathcal{R}_{i+1}n_{i+1}
    \end{equation}
    Substituting in the induction assumption we have the well known result:
    \begin{align}
    0=&-(\mathcal{J}_i+\cancel{\mathcal{R}_i})n_i+\cancel{\mathcal{R}_in_i}+\mathcal{R}_{i+1}n_{i+1}\rightarrow\\
    \rightarrow&\frac{n_i}{n_{i+1}}=\frac{\mathcal{R}_{i+1}}{\mathcal{J}_i}
    \end{align}
    This is easy to show for the first pair using \ref{eq:dotn0}=0. This recursive solution is quickly generalized for the relationship between $n_i$ and $n_j$, where $i<j$ and $i>j$ respectively:
    \begin{align}
        \nonumber
        n_i=&\frac{\mathcal{R}_{i+1}}{\mathcal{J}_i}n_{i+1}=\frac{\mathcal{R}_{i+1}}{\mathcal{J}_i}\frac{\mathcal{R}_{i+2}}{\mathcal{J}_{i+1}}n_{i+2}=...= \\
        =&n_j\prod_{k=i}^{j-1}\frac{\mathcal{R}_{k+1}}{\mathcal{J}_{k}}\\ \nonumber
        n_i=&\frac{\mathcal{J}_{i-1}}{\mathcal{R}_i}n_{i-1}=\frac{\mathcal{J}_{i-1}}{\mathcal{R}_i}\frac{\mathcal{J}_{i-2}}{\mathcal{R}_{i-1}}n_{i-2}=...= \\
        =&n_j\prod_{k=j}^{i-1}\frac{\mathcal{J}_{k}}{\mathcal{R}_{k+1}}
    \end{align}
    Finally we note that our system when summed is telescopic, that is:
    \begin{align}
    \sum_{i=0}^T \dot n_i=&0\rightarrow  \\
    N=&\sum n_i=n_i\left(1+\left(\sum_{j<i}+\sum_{j>i}\right)\frac{n_j}{n_i}\right)\label{eq:N}
    \end{align}
    where we have defined $N$ as the constant number of particles. Thus we obtain a complete closed form solution, starting from equation \ref{eq:N} and solving for $n_i$:
    \begin{align}
        n_i=N/\left(1+\sum_{j<i}\prod_{k=j}^{i-1}\frac{\mathcal{R}_{k+1}}{\mathcal{J}_{k}}+
        \sum_{j>i}\prod_{k=i}^{j-1}\frac{\mathcal{J}_{k}}{\mathcal{R}_{k+1}}\right)\label{eq:niss}
    \end{align}
    \end{appendix}
\end{document}